# Ultralow Thermal Conductivity and Large Figure of Merit in Low-Cost and Nontoxic Core-Shell Cu@Cu$_2$O Nanocomposites

Vikash Sharma[1,*], Gunadhor Singh Okram[1,*], Divya Verma[2], Niranjan Prasad Lalla[1], Yung-Kang Kuo[3]

[1]UGC-DAE Consortium for Scientific Research, University Campus, Khandwa Road, Indore 452001, Madhya Pradesh, India.

[2] Government College Alote, District Ratlam, 457114 Madhya Pradesh, India

[3]Department of Physics, National Dong-Hwa University, Hualien 97401, Taiwan

[*]Email: vikash@csr.res.in & okram@csr.res.in

**Abstract:** Identification of novel materials with enhanced thermoelectric (TE) performance is critical for advancing TE research. In this direction, this is the first report on TE properties of low-cost, nontoxic, and abundant core-shell Cu@Cu$_2$O nanocomposites (NCs) synthesized using a facile and cheap solution-phase method. They show ultralow thermal conductivity of nearly $10^{-3}$ of copper bulk value, large thermopower ~373 µVK$^{-1}$, and consequently, a TE figure of merit (ZT) of 0.16 at 320 K which is larger than those of many of the potential TE materials such as PbTe, SnSe and SiGe, showing its potential for TE applications. The ultralow thermal conductivity is mainly attributed to the multiscale phonon scattering from intrinsic defects in Cu$_2$O, grain boundaries (GBs), lattice-mismatched interface as well as dissimilar vibrational properties. The large thermopower is associated with sharp modulation in carrier density of states (DOS) due to charge transfer between Cu and Cu$_2$O nanoparticles (NPs), and carrier energy filtering.

**Keywords:** Nanocomposites; Interface; Defects; Ultralow thermal conductivity; Figure of merit

## 1. Introduction

Thermoelectricity as a conversion of heat into electricity directly acquires a significant attention for a sustainable green energy solution to the energy crisis of the world of inevitable fossil fuel depletion and environmental challenges[1,2,3]. The conversion efficiency of TE material or device can be characterized by $ZT = \frac{S^2 \sigma}{\kappa_e + \kappa_l} T$, where S is the Seebeck coefficient, $\sigma$ is the electrical conductivity, $\kappa = \kappa_e + \kappa_l$ is the total thermal conductivity with $\kappa_e$ and $\kappa_l$ as electronic and lattice contribution to $\kappa$ and T is the absolute temperature[1]. Notably, large ZT and PF ($S^2\sigma$) for high conversion efficiency and power output respectively are essential ingredients of competitive TE materials. The main obstacle to achieve high ZT in the usual materials has been inherent conflict among the transport parameters S, σ, and κ. Considerable efforts have been attempted through various approaches including nanoinclusion of metallic nanoparticles in semiconducting host[2,3,4,5], NCs[6,7], and metal/semiconductor interfaces[8] especially in the classically high-performance chalcogenides such as $Bi_2Te_3$, SnSe and PbTe over many decades for the enhancement of ZT. However, not only that they get oxidized easily at higher temperature, but also their constituent elements are costly, toxic and scarce[3,9]. To overcome these problems, researchers are devoting their significant efforts to identify the oxide-based TE materials[3,9].

So far, good oxide TE materials are p-type conduction $NaCo_2O_4$, $Na_xCoO_2$ and $Ca_3Co_4O_{9+\delta}$ and n-type conduction $SrTiO_3$, $CaMnO_3$ and ZnO[9]. In this context, significantly large value of S in p-type cuprous oxide $Cu_2O$ with direct band gap (~ 2.17 eV)[10] and melting point ~1508 K has been reported in experimentally[11,12,13,14,15] and theoretically[16,17]. For instance, Young et al.[11] reported S around 1050 µV/K at 500 K for single crystalline $Cu_2O$. Figueira et al.[12] reported 1000 µV/K at 300 K with hole concentration of $4\times10^{16}$ cm$^{-3}$. Hartung et al.[13] observed S ~900 µV/K at 300 K. Linnera et al.[16] predicted S between 870 µV/K to 970 µV/K for high and low hole concentration and S ~ 1200 µV/K at 300 K with hole concentration of $3\times10^{15}$ cm$^{-3}$. Andrei et al.[14,15] showed S in the range of 600 to 650 µV/K at 298 K. Chen et al.[17] calculated theoretically S of 500 µV/K at 600 K and 200 µV/K at 500 K with hole concentration of $1 \times 10^{19}$ cm$^{-3}$ and $5.2 \times 10^{20}$ cm$^{-3}$, respectively. In spite of large value of S, its TE properties have been less interesting due to its low electrical conductivity, and hence their experimental study is limited only to determine the sign of charge carriers. Significant works are therefore essential to improve the performance of $Cu_2O$ as a TE material[18].

This scenario motivated us for investigations of TE properties of nontoxic, low-cost and abundant $Cu/Cu_2O$ NCs with large value of S of semiconducting $Cu_2O$[11,12,13,16,14,15,17] along with one of the best electrically conducting elements, Cu. We show the ultralow value of κ ~ 0.49 Wm$^{-1}$K$^{-1}$ and ZT ~ 0.16 at 320 K in the reasonably well-characterized core-shell Cu@$Cu_2O$ NCs synthesized using a facile and environmental-friendly solution phase method. This value of ZT is significantly larger than many of the potential TE materials based-on PbTe, SnSe and SiGe showing potential for TE applications[1,19]. With the sign of thermal stability up to 700 °C seen, value of κ lower than that of silicon, indicating application potential in NC-based heat dissipation in electronics and photonics[20]. The possible origins of these significant results have been discussed.

## 2. Experimental
### 2.1 *Synthesis of core-shell Cu@Cu$_2$O nanocomposites*

The NCs were synthesized using the as-received copper-acetate monohydrate, CAM (99.99%), trioctylphosphine (TOP, 90%) and oleylamine (OA, 70%) from Sigma Aldrich. Typically, preheated 0.25 ml TOP at 200 °C was injected in the already degassed mixture of 8 mmol of CAM in 8 ml of OA at 120 °C for 30 min, and was heated at 190 °C for 2 h under nitrogen atmosphere. We cooled down the product to room temperature, washed it in n-hexane, and separated via centrifugation at 12000 rpm for 5 minutes. This was repeated four times to remove the access OA, TOP, and acetate, and the product was dried at 60 $^0$C to obtain the NC powder. The code for this sample is NC1. Samples prepared in 1 ml, 3 ml, 5 ml, 8 ml and 10 ml TOP, respectively, with other conditions remaining the same are coded as NC2, NC3, NC4, NC5 and NC6. This is schematically illustrated in figure 1, and table S1 summarizes these details.

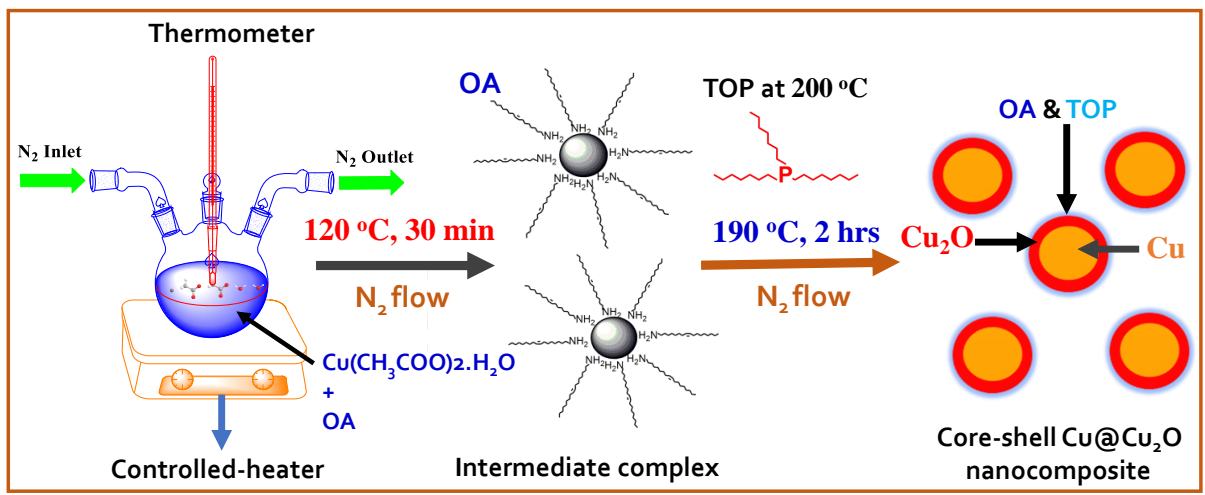

**Figure 1** Schematic representation of synthesis of core-shell Cu@Cu$_2$O nanocomposite using copper acetate hydrate, oleylamine (OA) and trioctylphosphine (TOP).

## 2.2 *Ligand removable process*

The electrical conductivity reducing insulating ligands OA and TOP because of decease of mobility of charge carriers, their scattering, high interfacial densities and impurities were removed without considerable modification in size and morphology of nanoparticles as described in ref.[21]. We annealed cold-pressed pellets of NC6 at different temperatures from 300 °C to 600 °C for different time intervals of 10 min to 180 min under high vacuum ~$10^{-8}$ torr. Fourier transformed infrared (FTIR) measurements were performed on the fine powder made by crushing the pellets to see the presence of vibrational modes of OA and TOP.

## 2.3 *Experimental techniques*

X-ray diffractometer (Bruker D8 Advance) with Cu K$_\alpha$ and spectrometer Bruker Vertex 70 were used for collecting X-ray diffraction (XRD) patterns and Fourier transformed infrared (FTIR) spectra, respectively. Transmission electron microscopy (TEM) and field emission scanning electron microscopy (FESEM) measurements were performed using TECHNAI-20-G$^2$ (200 KV) and FEI Nova nanosem450, respectively. Scanning electron microscope (SEM) equipped with EDX JEOL JSM 5600 was used for energy dispersive X-ray (EDX) measurements. Raman measurements were performed at room temperature using a Jobin Yvon Horiba LABRAM-HR Visible instrument equipped with a semiconductor diode laser of wavelength 633 nm. The resistance measurements were performed using four-point probes and Seebeck coefficient measurements using differential direct current setup in the temperature range of 5–320 K in a specially designed commercially available Dewar[22]. Thermal conductivity was measured in the temperature range of 10 - 320 K using a dc pulse laser technique[23]. The EDX, Raman and transport measurements were performed on cold-pressed pellets from ligand-removed and annealed powder samples. Pellets were made from the powder in a rectangular die of 8 mm×3 mm size with a uniaxial pressure of ~ 2.5 GPa. The mass densities of pellets are found to be around 8.0 g/cm$^3$, 7.8 g/cm$^3$, 7.7 g/cm$^3$, 7.5 g/cm$^3$, 7.3 g/cm$^3$ and 7.0 g/cm$^3$ for NC1, NC2, NC3, NC4, NC5 and NC6, respectively. The error in the estimation was about 5%. The average mass densities were calculated to be around 8.8 g/cm$^3$, 8.6 g/cm$^3$, 8.5 g/cm$^3$, 8.3 g/cm$^3$, 8.1 g/cm$^3$ and 7.8 g/cm$^3$, respectively using $\rho = f_{Cu} * \rho_{Cu} + f_{Cu2O} * \rho_{Cu2O}$, where $f_{Cu}$, and $f_{Cu2O}$ are volume phase fractions and $\rho_{Cu}$, and $\rho_{Cu2O}$, mass densities of Cu and Cu$_2$O, with $\rho_{Cu}$ = 8.9 g/cm$^3$ and $\rho_{Cu2O}$ = 6.0 g/cm$^3$, $f_{Cu}$ and $f_{Cu2O}$ were from Rietveld refinement of XRD of these nanocomposites (table S2).

## 3. Results and discussion
### *3.1 Structural and microscopic study*

Analysis of XRD patterns of NC1, NC2, NC3, NC4, NC5 and NC6 indicates simultaneous presence of fcc phases of both Cu (JCPDS# 851326) and Cu$_2$O (JCPDS#782076) at varying fractions (figure 2a and table S2); crystallite size obtained from Scherrer formula for both Cu and Cu$_2$O phases are listed in table S1. As quantity of TOP increases, crystallite size of both Cu and Cu$_2$O increases from NC1 to NC3 and then decreases from NC4 to NC6, and the ratio of crystallite size of Cu to Cu$_2$O (Cu$_{Size}$:Cu$_2$O$_{Size}$) shows the reverse trend compared to individual crystallite size (figure 2a, inset and table S1). Figure 2b shows XRD of annealed NC6 at different temperatures of 300-600 °C and time interval of 10-120 min to check the stability at high temperature. As T increases at 300 °C and 400 °C, XRD peak broadening does not considerably change with annealing for 10 min while increase in time to 120 min leads to narrowing the peaks suggesting increase in crystallite size.

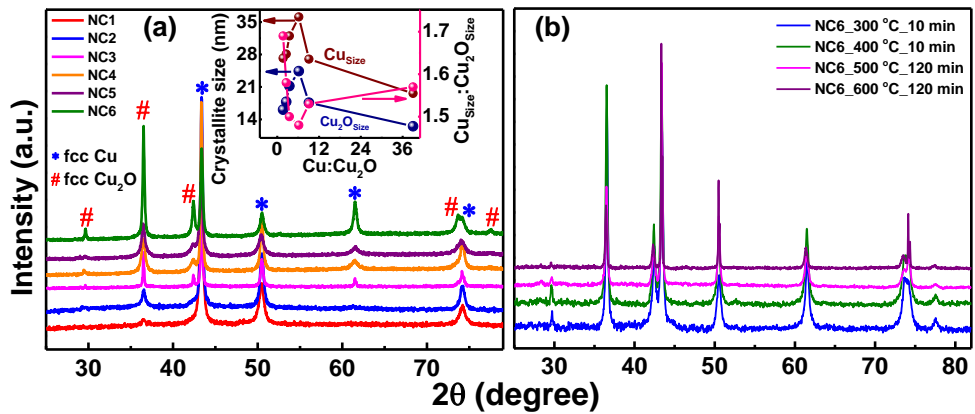

**Figure 2** (a) X-ray diffraction patterns of NC1 to NC6. Symbols # and * identify the fcc phases of Cu$_2$O and Cu, respectively. Inset: crystallite size of Cu and Cu$_2$O, and their size ratio Cu$_{Size}$:Cu$_2$O$_{Size}$ versus their volume phase fraction ratio (Cu:Cu$_2$O). (b) X-ray diffraction patterns of NC6 annealed at different temperatures and time durations.

Further, increase in annealing temperature to 500 °C and 600 °C for 120 min, effectively increases the crystallite size (table S1), manifesting agglomeration of particles or crystal growth.

The NC6 annealed at 400 °C for 10 min shows the nearly equal crystallite size and volume phase fractions for both Cu and $Cu_2O$, which is consistent with earlier report[24]. Therefore, we annealed NC1 to NC6 at 400 °C for 10 min for removing OA and TOP with particular aim to improve their electrical properties; removal of OA and TOP has been confirmed using FTIR as discussed later. Rietveld refinements of XRD data of annealed NC samples at 400 °C for 10 min were done using Fullprof_suite software (figure 3a-f) to get better information about lattice parameters and phase fractions of Cu and $Cu_2O$, and obtained parameters are listed in table S2. The volume phase fraction of Cu to $Cu_2O$ (Cu:$Cu_2O$) decreases in NC1, NC2, NC3, NC4, NC5 and NC6 (table S2) but the lattice parameters for $Cu_2O$ and Cu decrease from NC1 to NC3 and from NC4 to NC6. The lattice mismatch is defined as $\frac{a_{Cu2O}-a_{Cu}}{a_{Cu2O}}$, where $a_{cu}$ and $a_{cu2O}$ are respective lattice parameters of fcc Cu and fcc $Cu_2O$. It is around 15.3 % in these NCs (table S2) that is excellently consistent with reported value of ~15.4% in their bulk form[25].

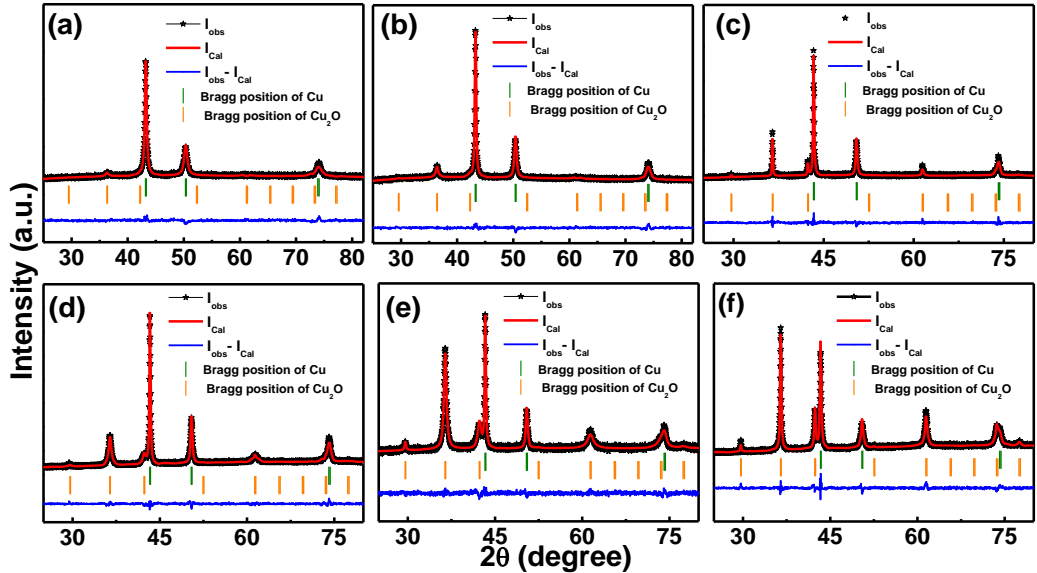

**Figure 3** Rietveld refinements of (a) NC1, (b) NC2, (c) NC3, (d) NC4, (e) NC5 and (f) NC6. Vertical lines indicate the Bragg peak positions of the chemical phases of Cu and $Cu_2O$, and blue curves are difference between the observed and calculated curves.

Figure 4a shows the FTIR spectra of pristine NC6, and after annealing it at 300 °C and 400 °C for 10 min under high vacuum ~$10^{-8}$ torr in order to confirm removal of OA and TOP or not from NCs' surface after annealing. A sharp mode near 628 $cm^{-1}$ due to Cu-O stretching of $Cu_2O$[26] is survived up to 400 °C. A broad mode near 3465 $cm^{-1}$ is due to O-H stretching of $H_2O$ absorbed from air. Dips due to bending vibration (δ (C=H)), asymmetric stretching ($\nu_{as}$ (C-H)) and symmetric stretching ($\nu_s$(C-H)) mode of $CH_2$ are seen in NC6 near 2964, 2927 and 2856 $cm^{-1}$, respectively[27]. The bending vibrations δ($NH_2$) and δ(-C=C) of $NH_2$ appear near 1631 $cm^{-1}$ and 1595 $cm^{-1}$, respectively, and a dip near 1382 $cm^{-1}$ is assigned to $CH_2$ wagging vibration (ω($CH_2$))[27].

Vibrations at 1058 cm$^{-1}$ and 728 cm$^{-1}$ are also seen due to bending vibrations of C-N ($\delta$ (C-N)) and -C-C ($\delta$(-C-C)) of OA and TOP[27]. These results indicate presence of OA and TOP on the surface of NC6. However, we can clearly see suppression of these vibrational modes for NC6 heated at 300 °C, but nearly nil at 400 °C. The feeble intensity of a few modes may suggest their reduced presence. Therefore, effect of OA and TOP are suppressed significantly at 400 °C for 10 min without considerable modification in the size and morphology of these NCs, which is in line with previous report[21].

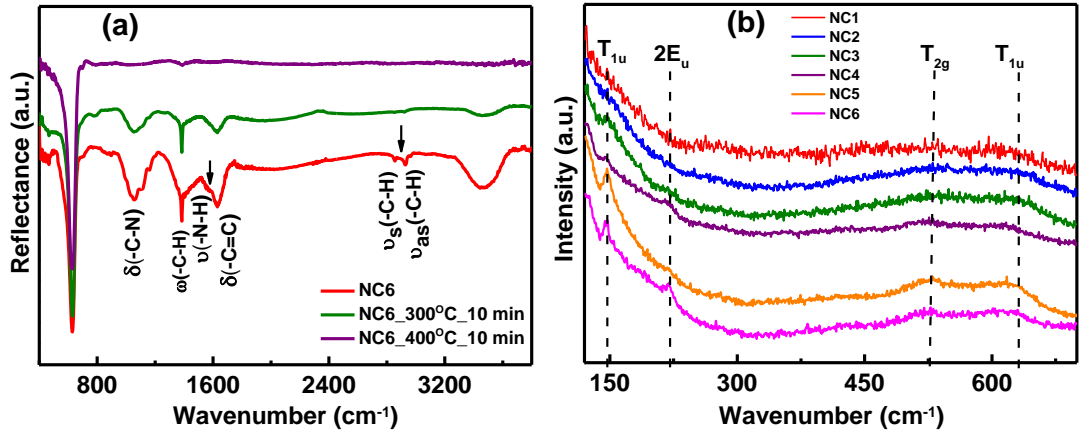

**Figure 4** (a) Fourier-transformed infrared spectra of pristine NC6, and after annealing at 300 °C and 400 °C for 10 min and (b) Raman spectra of NC1 to NC6.

Intrinsic point defects may appear in $Cu_2O$. They may be oxygen vacancies $V_O$, and copper vacancies $V_{Cu}$[10,28]. $V_O$ may be of two types of oxygen interstitials $O_{i,oct}$ with six a neighbors nd $O_{i,tetr}$ with four Cu neighbors. $V_{Cu}$ may be antisite defects viz, $Cu_O$ and $O_{Cu}$, and copper vacancies in split configuration $V_{Cu}^{split}$ in which a Cu atom adjacent to the vacancy moves into a position between the corresponding two Cu lattice sites. It is well-known that $V_{Cu}$ and $V_{Cu}^{split}$ are the most stable vacancies owing to their relatively lower energy of formation, and lead to the p-type conduction in $Cu_2O$[10,28]. We, therefore, employed the Raman spectroscopy in order to see these defects in $Cu/Cu_2O$ NCs. Only one one-phonon mode $T_{2g}$ should appear in Raman spectrum of perfect crystal of $Cu_2O$[28]. It is reported that some other infrared active and silent lattice modes may become Raman allowed due to the reduction of symmetry because of the point defects in $Cu_2O$[28]. Raman spectra of NC1 to NC6 are shown in figure 4b. The vibrational modes near 146 cm$^{-1}$, 221 cm$^{-1}$, 528 cm$^{-1}$ and 628 cm$^{-1}$ in NC6 are assigned to the $T_{1u}$ (TO i.e. transverse optical), $2E_u$, $T_{2g}$ and $T_{1u}$ (TO), in line with earlier report[28]. The two $T_{1u}$ modes are due to infrared active optical lattice vibrations, $2E_u$ is two-phonon mode that is well-defined silent lattice mode, and $T_{2g}$ mode is threefold-degenerate Raman active mode[28]. The mode $T_{2g}$ is found to be near, 530 cm$^{-1}$ in NC3 to NC5, and 532 cm$^{-1}$ and 534 cm$^{-1}$ in NC1 and NC2, respectively while other modes appeared at nearly the same position in NC1 to NC5 as NC6. Notably, the appearance of the infrared and silent lattice modes $T_{1u}$, $2E_u$ and $T_{1u}$ confirms the point defects $V_{Cu}^{split}$ in $Cu_2O$[28]. Furthermore, intensity of these modes increases as $Cu:Cu_2O$ decreases, pointing to increases in degree of defects.

Interestingly, as Cu:Cu$_2$O increases, T$_{2g}$ mode shifts towards lower wavenumber side (figure 4b) indicates the softening of Raman active mode T$_{2g}$. These point defects and softening in T$_{2g}$ mode lead to the reduced lattice thermal conductivity, $\kappa_l$, discussed later.

Figure 5a-c shows the representative TEM micrographs of NC3, NC5, and NC6. Clearly, two types of NPs with different colour contrast are seen. The black portions in the inner region with cuboid shapes appear most probably due to Cu while that one seen at the boundaries as Cu$_2$O shell to form core-shell-like Cu@Cu$_2$O NCs, in line with earlier report[29]. Particle size ranging from 30 nm to 50 nm for NC3 (figure 5a), and 32 nm to 49 nm for NC5 (figure 5b) with average particle size around 40 ± 3 nm and 35 ± 4 nm, respectively, are seen.

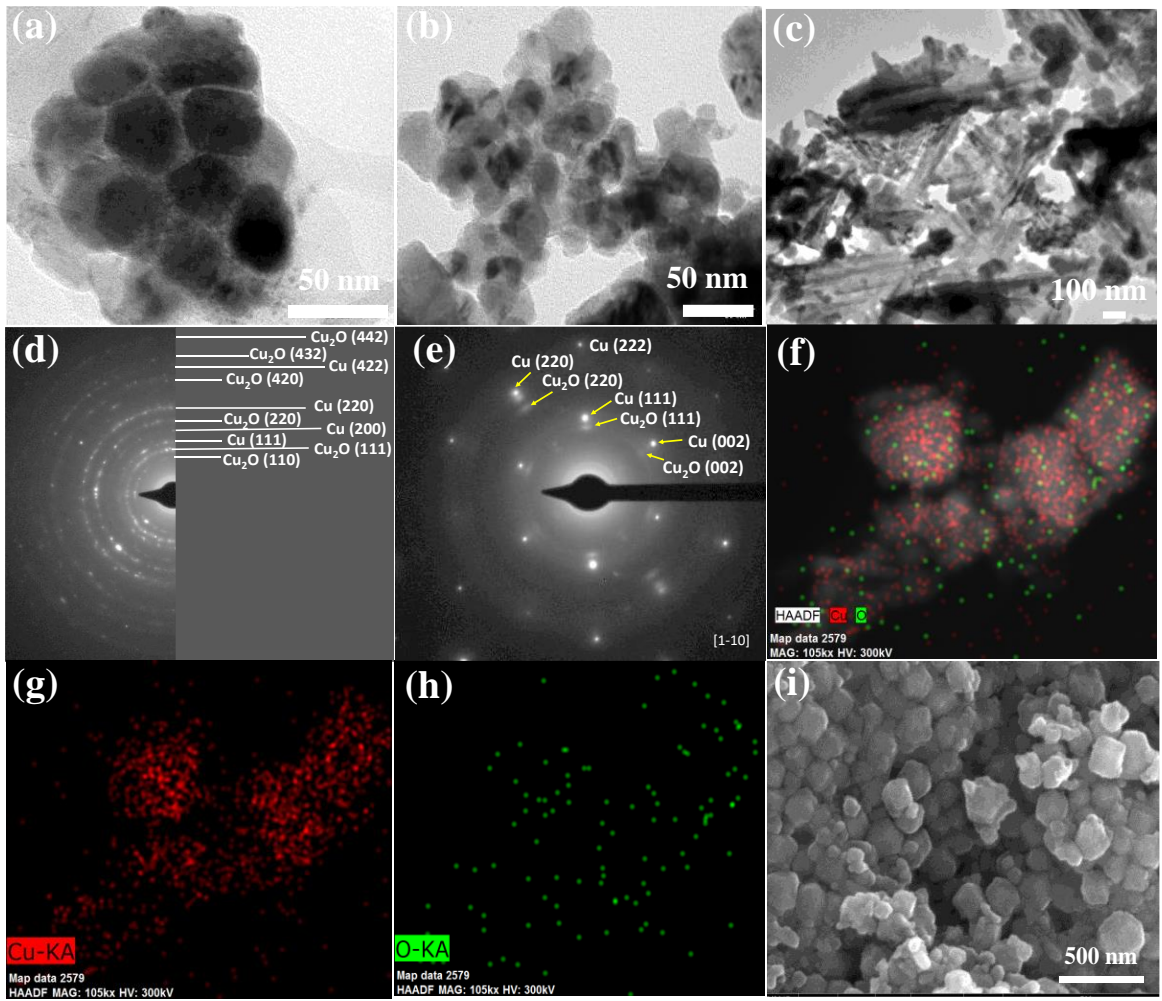

**Figure 5** Transmission electron micrographs of (a) NC3, (b) NC5 and (c) NC6, and SAD patterns of (d) NC3 and (e) NC6. Energy dispersive X-ray spectroscopic (EDX) mapping of combined Cu and O (f), separate Cu (g) and O (h), and (i) field emission scanning electron microscopy (FESEM) micrograph of NC3.

However, NC6 shows (figure 5c) particles with different morphologies of rods and cuboids with average diameter of rods and cuboids around 25 nm and 45 nm, respectively. The polycrystalline nature of NC3 is evident from selected area diffraction (SAD) pattern (figure 5d), which shows the reflections from (110), (111), (220), (420), (432) and (442) planes of $Cu_2O$, as well as (111), (200), (220) and (442) planes of Cu. Their particular sequence is also evident in NC6 (figure 5e), wherein each Cu reflection follows closely that of $Cu_2O$. The relatively more defused $Cu_2O$ reflections compared to that of Cu is indicative of an epitaxial growth of $Cu_2O$ layers on large Cu crystal nanoparticles, i.e. it clearly shows the formation of core-shell $Cu@Cu_2O$ structure. This is further proven in energy dispersive X-ray spectroscopy (EDX) mapping in NC3 (figure 5f-h) that clearly shows the presence of only Cu and O with the presence of more concentration of Cu in the central region and less concentration of O (i.e. $Cu_2O$) at the periphery. This proves further the formation core-shell $Cu@Cu_2O$ structure, which is also evident from figure S1a. Field emission scanning electron microscopy (FESEM) image of NC3 also confirms the cuboid-like clusters (figure 5i).

The EDX spectrum of NC6 exhibits peaks of Cu and O only (figure S1b). It indicates the absence of other impurities. The atomic percentage of Cu and O found is 66.6 % and 33.4 %, respectively. It shows much larger fraction of Cu compared to O in NC6. When we consider the formation of stoichiometric $Cu_2O$, 33.4 % of O requires 66.8 % of Cu in the NC6 $Cu@Cu_2O$ composite. This is in contrast to 63.5% volume fraction of Cu and 36.5 % volume fraction of $Cu_2O$ from XRD analysis (table S2). As per this XRD analytical data of 36.5 % volume fraction of $Cu_2O$, O content comes out to be about 12.2 %, which is much smaller than that seen from EDX i.e. 33.4 %. This may indicate the presence of O-excess $Cu_2O$ in consistent with Raman analysis (figure 4b) and probably extra adventitious O in the composite, deviating significantly from its stoichiometry. This in turn will imply higher electrical conductivity of $Cu_2O$ and hence p-type conduction in this NC due to the excess O; this will be discussed later in electrical transport studies. Also considering that volume fraction of Cu NPs is rather high; oxidation occurs as the reaction progresses. This process gradually increases as we increase the concentration of TOP for the same duration of 2 h sample preparation. As a result, $Cu_2O$ coats the Cu NPs, in line with the results of TEM (figure 2a,b) and EDS (figure 2e-g) data. This is in line with how the crystallite size increases initially and then decreases as the concentration of TOP increases in these components (Table S1). In fact, Cu (111) facet has lower work function and better stability by means of surface coordination function compared to others. Consequently, an intrinsic $Cu(111)/Cu_2O(111)$ interface in the form of core-shell $Cu@Cu_2O$ is formed[30]. These kind of core-shell structures are highly essential to obtain better thermoelectric properties of these NCs.

### 3.2 Thermoelectric Properties
#### *3.2.1 Electrical resistivity and Seebeck coefficient*

The evolution of resistivity ρ from NC1 to NC6 as a function of T over 5 K to 330 K as concentration of Cu decreases is shown in figure 6a-d. Clearly, NC1 to NC3 exhibit metallic-like resistivity behaviour with increasing trend of slopes 1.04, 3.14 and 5.24 nΩm/K with enhancing resistivity as the fraction of Cu decreases. This scenario has however completely changed starting from NC4 wherein we observe metallic-like behaviour below 244 K and semiconducting above with a rounded peak-like feature in the resistivity range of 40.5-49.5 µΩm. A much broader rounded peak-like feature near 105 K with resistivity range of 0.170-0.213 mΩm in NC5 is seen.

Enhanced trend is seen in NC5 with its metallic-like behaviour below 105 K and semiconducting above this. Similar trend holds true in NC6 with a still much broader rounded peak-like feature near 85 K with resistivity range of 0.538-0.610 mΩm with its metallic-like behaviour below 85 K and semiconducting above this. This trend of metal to semiconductor transition (MST) in resistivity as concentration of Cu decreases (table S2) is rather interesting. This is comparable with earlier report in Cu/$Cu_2O$ composites obtained from oxidation of Cu[31] wherein MST starts near 110 K, but did not cover the full range of transitions due to their experimental limitations and for Ag/$Ag_2S$ NCs[5]. Therefore, metallic behaviour with high fraction of Cu in NC1, NC2 and NC3 corroborates the core-shell Cu@$Cu_2O$ structure and EDS data above (figure 5) such that core Cu NPs are oxidized to form $Cu_2O$ shells in such way that the role of $Cu_2O$ is quite marginal in these NCs. However, in NC4, NC5 and NC6, role of $Cu_2O$ gradually becomes much more dominating over Cu in terms of magnitude and behaviour of resistivity as $Cu_2O$ as shell gradually become thicker and Cu as core thinner, as concentration of TOP increases i.e., as concentration of Cu decreases (table S2). In this scenario, switching to MST transition from metallic behaviour is well explained as both the components turning into smaller size particles (table S1) and hence more resistive. Increase in scattering of electrons with the point defects in $Cu_2O$, interface, and interactions of Cu and $Cu_2O$ is attributed to the origin of MST. As the concentration of O-excess $Cu_2O$ increases, its semiconducting behaviour gradually sets in, in addition to the metallic nature of Cu. This is in line with relatively low values of the resistivity even for the maximum $Cu_2O$ content. The different degrees of band bending produced by different facets of Cu and $Cu_2O$ can significantly affect the electrical properties, as $Cu_2O$ (111), (100), and (110) surfaces have respective metallic, semi-metallic, and semiconducting band structures as reported in W/$Cu_2O$[32,33]. Their electrical transport in metal/ semiconductor interface will depend on the particle size of Cu and $Cu_2O$, their phase fraction, formed interface facets and contact area or bonding strength[10,30].

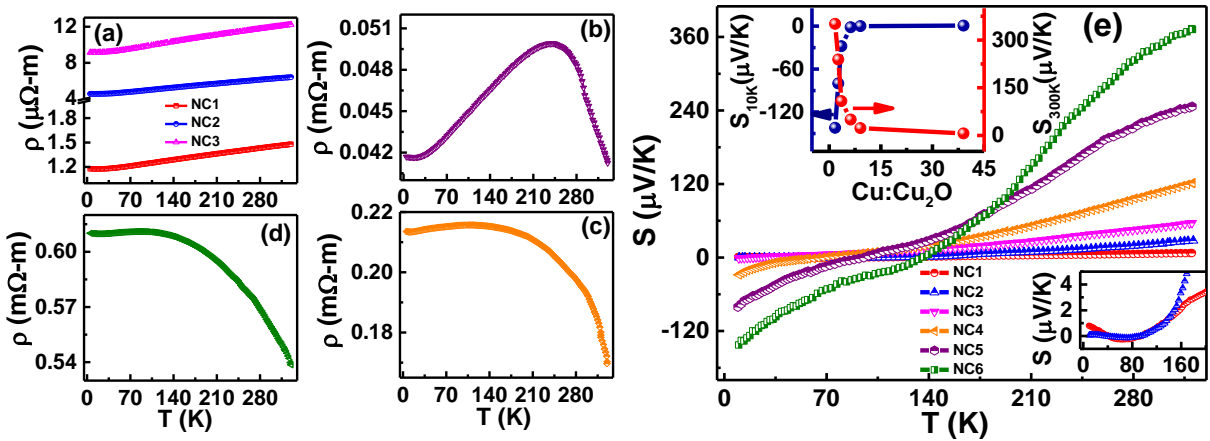

**Figure 6** Resistivity ρ of (a) NC1, NC2 and NC3, (b) NC4, (c) NC5 and (d) NC6. (e) thermopower S of NC1 to NC6. Insets: (e), top, thermopower at 10 K and 300 K versus Cu concentration, bottom, expanded view of NC1 and NC2.

Figure 6e shows the evolution of Seebeck coefficient S in NC1 to NC6, as concentration of $Cu_2O$ increases (table S2). As T increases, S of NC1 decreases up to 68 K (figure 6e, bottom inset), then starts to rise above it. We attribute this minimum near 68 K with value -0.24 µV/K to phonon drag minimum (PDM). Interestingly, S of NC1 exhibits positive value in the T range 96 K to 330 K. Unlike the phonon drag peak for bulk Cu, Seebeck coefficient of Cu NPs or NC1 shows a valley-like feature and crossovers from positive to negative. This indicates that size effect modifies the TE transport as it depends upon the slope of $\sigma(E)$ or $\tau(E)$. It changes with particle size and temperature[27]. There are two crossovers of S in NC1 and NC2 with its sign change from positive to negative below 43 K and 34 K, and again negative to positive above 96 K and 98 K, respectively. PDM is significantly supressed in NC2, and completely wiped out in NC3. The positive value of S below 43 K and 34 K of NC1 and NC2 turns negative below 19 K, 66 K, and 94 K of NC3, NC4, and NC5, respectively. The suppression in PDM with decrease in Cu:$Cu_2O$ ratio is in line with decrease in electron-phonon coupling strength. We expect this in metal having defects and formed interface with oxide. Particularly, Cu@$Cu_2O$ interface (figure 5) significantly affects the transmission of electrons and phonons, and increases their scattering with defects (vacancies, dislocations, point defects, and GBs) in $Cu_2O$ or Cu. The absolute value of S at 10 K and 300 K increases very fast as Cu concentration decreases that turns out to be exponential (figure 6e, top inset). We attribute the positive and negative signs of S with increase in $Cu_2O$ fraction to presence of both types of charge carriers due to core-shell Cu@$Cu_2O$ interfaces and associated inherent defects. Also sign of S for metals or degenerate semiconductor is decided by slope of energy-dependent electrical conductivity $\sigma(E)$ or relaxation time $\tau(E)$, which can change with T and any modification in DOS with composition or size[1,34].

*3.2.2 Possible mechanism of large Seebeck coefficient*

The individual work functions (figure 7a) of Cu and $Cu_2O$ may be possible to modify the band diagram for ideal Cu@$Cu_2O$ composite (figure 7b). When Cu metal is in contact with p-type semiconductor $Cu_2O$,[32] valence and conduction band edges near the interface is likely to bend towards the metal (figure b) [25]. In the equilibrium state without an external field, their Fermi levels are equalized with proper band bending, which affects both the space charge near the interface, and diffusive transport of charge carriers across the interface[25]. This introduces a potential barrier ($V_B$) ~ 0.74 eV for holes[35] due to their different work functions (figure 7a)[30]. Subsequently, Fermi level pinning by DOS at the interface, metal-induced gap states and defects associated states may appear[8]. $V_B$ will scatter low energy carriers while higher energy carriers than the potential barrier will be passed through (figure 7b). This filtration of low energy charge carriers increases S[3,4,6]. Notably, the degree of band bending will be highly dependent on bonding between Cu and $Cu_2O$, surface coverage of $Cu_2O$ i.e. it is partially covered or fully covered with variation in contact area[36]. We expect its maximum in NC6 with larger phase fraction of $Cu_2O$. In fact, figure 7 considers the ideal situation without any defect. This however is not so practically since NCs contain various types of defects including point defects, dislocations, vacancies, trap states and GBs, in consistent with earlier reports[10,30,33,36].

The Seebeck coefficient S can be given by Mott formula as[34]

$$S = -\frac{\pi^2 k_B^2 T}{3e} \left[ \frac{dn(E)}{ndE} + \frac{1}{\mu} \frac{d\mu(E)}{dE} \right]_{E=E_F}, \quad (1)$$

Where $k_B$, e, $n(E)$ and $\mu(E)$ are Boltzmann constant, elementary charge, energy-dependent carrier concentration and mobility, respectively.

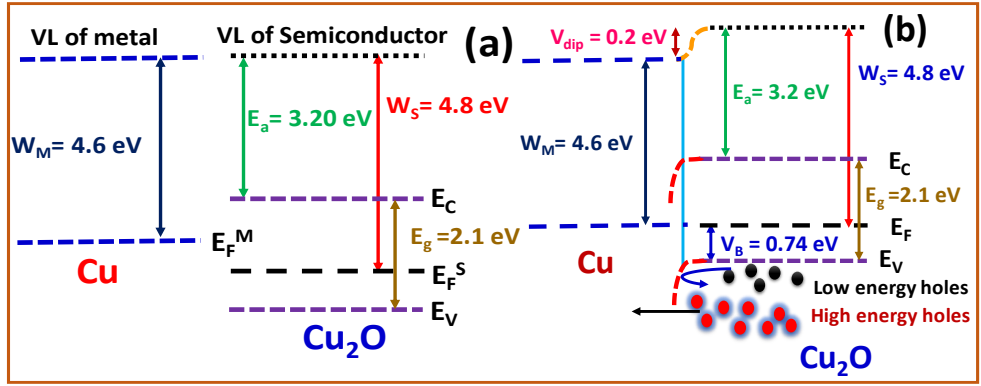

**Figure 7** (a) Modified band diagram of $Cu_2O$ and Cu before contact and (b) in contact in equilibrium; VL, vacuum level. The known energy level of defect less Cu and $Cu_2O$ are used. $W_M$, $W_S$, $V_B$, $E_a$, and $V_{dip}$ are metal work function, semiconductor work function, potential barrier of Cu and $Cu_2O$ contact, electron affinity of $Cu_2O$ and potential of formed interfacial dipole, respectively. $E_F^M$, $E_F^S$, $E_V$, $E_c$, $E_T$, and $E_F$ are Fermi level of metal, Fermi level of semiconductor, valence band energy, conduction band energy, transport energy level, and Fermi level.

From eq. 1, it is clear that sharp features in the carrier density of states (DOS) around $E_F$ i.e. increase in $\frac{dn(E)}{dE}$ and energy filtering effect that selectively scatters low energy charge carriers i.e. higher $\frac{d\mu(E)}{dE}$ result in larger value of S[34,37]. The sharp features in DOS due to size and defects and vacancies in $Cu_2O$ likely to enhance the S[34,38]. Charge can be transferred from Cu to $Cu_2O$ in $Cu/Cu_2O$ with $Cu_2O$ gaining ~ 2.6 electrons that may lead to formation of new bonds between the $Cu_2O$ and Cu surface, and hence DOS near the Fermi level enhances[30]. This is favourable for increase in S[30]. The energy filtering effect for $Cu/Cu_2O$ interface is shown in figure 7 that leads to increase in S by increase the second term in eq. 1. Thus, both terms in eq. 1 are effectively involved in the enhancement in S in these NCs[34,30,38].

### 3.2.3 Thermal conductivity

Figure 8a depicts evolution of thermal conductivity κ from NC1 to NC6. As T decreases, κ slowly reduces but steeper below ~ 130 K. It decreases initially monotonically faster with decrease in T, especially at low T, with different slopes and magnitudes as $Cu:Cu_2O$ ratio decreases. The κ increases with T even above 330 K. A broad hump around 110 K in NC1 shifts towards lower T in NC2-NC6. These trends can be understood in the following ways. Thermal transport across the metal/semiconductor interface involves both electrons in metals and phonons in semiconductors as they are the primary heat carriers. The transport of heat may be governed by three types of interactions: (i) phonons of metal with phonons of semiconductor (ii) electrons and phonons of metal and phonons of semiconductor and (iii) electrons of metal with phonons of semiconductor directly.

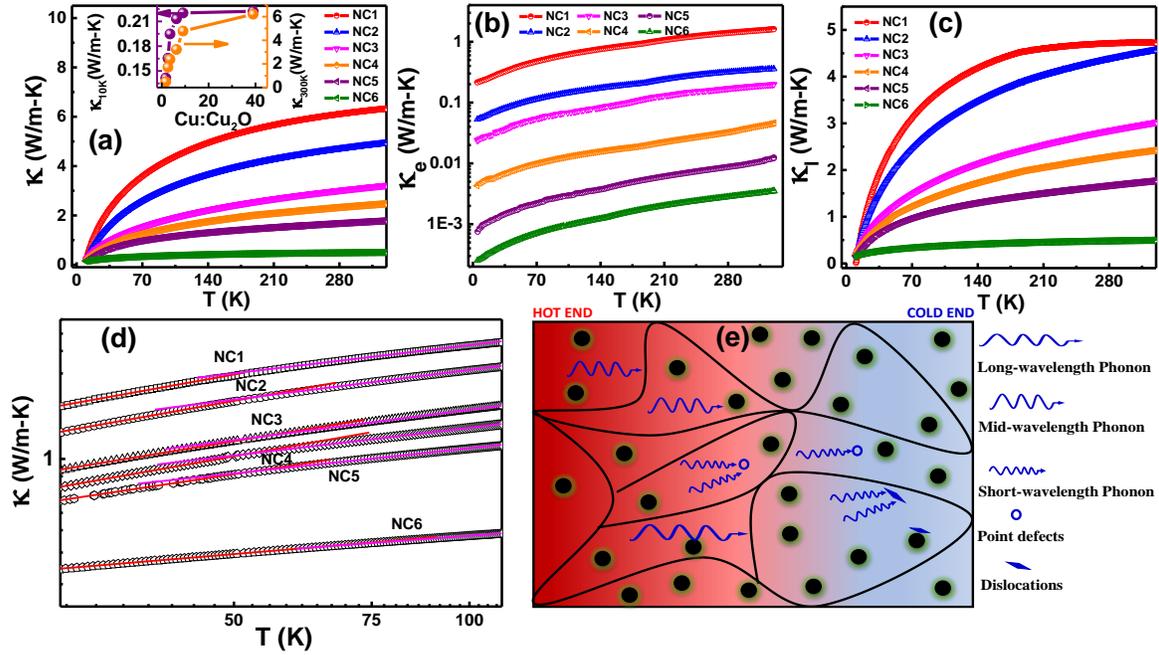

**Figure 8** (a) Total thermal conductivity, (b) electronic thermal conductivity, (c) lattice thermal conductivity of NC1 to NC6, and (d) fitted thermal conductivity. Inset shows its value at 10 K and 300 K as a function of Cu:Cu$_2$O. (e) Schematic representation of phonon scattering process.

These mechanisms depend on the type of interfacial barrier, lattice mismatch, point defects, dislocation, density of heat carrier near the interface and interfacial bonding between metal and semiconductor[39,40,41]. Therefore, these trends in κ may suggest involvement of interactions (i) and (iii)[42], and are somewhat similar to those in metal/oxide TiN/MgO composities[39], wherein the diffuse-mismatch model (DMM) with acoustic phonons was invoked without considering electron-phonon interaction (ii). Thus, monotonic decrease in κ near 300 K cannot be explained by using elastic phonon-phonon interaction (i) alone. The thermal conductivity κ at 10 K and 300 K decreases with decrease in Cu:Cu$_2$O or increase in Cu$_2$O fraction (figure 8a, inset). They strongly suggest the systematic increase in scattering of electrons and phonons with interface, defects and GBs as Cu$_2$O fraction increases. Similar trends have been calculated theoretically for CoSi$_2$(111)/Si(111) and Mo$_{1-x}$W$_x$S$_2$ alloy embedded with triangular WS$_2$ nanodomains[43,42], and experimentally in Si nanowires[44].

In order to understand better, $\kappa_e$ and $\kappa_l$ were calculated separately. The $\kappa_l$ is calculated using the relation $\kappa_l = \kappa - \kappa_e = \kappa - L\sigma T$, in order to assessed the influence of defects and interface on $\kappa_l$, where L is Lorentz number of 2.45×10$^{-8}$ WΩ-K$^{-2}$ for degenerate limit[45]. This, however, decreases for non-degenerate semiconductors which can be calculated by an empirical formula[45].

$$L \times 10^{-8} = 1.5 + \exp[-\frac{|S|}{116}], \qquad (2)$$

where L is in $W\Omega\text{-}K^{-2}$ and S in $\mu V/K$. We have calculated L using eq. 2 by considering the Cu@Cu$_2$O NC as non-degenerate semiconductor since Cu$_2$O is also a non-degenerate semiconductor. The Lorentz number L is $2.44\times10^{-8}$, $2.3\times10^{-8}$, $2.15\times10^{-8}$, $1.84\times10^{-8}$, $1.63\times10^{-8}$ and $1.55\times10^{-8}$ $W\Omega\text{-}K^{-2}$ for NC1, NC2, NC3, NC4, NC5 and NC6, respectively. The value of L for NC1 excellently matches exactly with degenerate limit. L decreases as Cu$_2$O increases, and is 36 % less for NC6 compared to $2.45\times10^{-8}$ $W\Omega\text{-}K^{-2}$. The latter is exactly the same as in acoustic phonon scattering case[45]. Thus, decrease in L with increase in Cu:Cu$_2$O is correlated with increase in defects and S[45].

The $\kappa_e$ increases linearly with temperature in all NC1 to NC6 above about 70 K but decreases faster below it (figure 8b). This shows that scattering of electrons increases while the number of phonons reduces for thermal conductance across the interface at low temperature. As fraction of Cu$_2$O increases, $\kappa_e$ decreases that is expected due to stronger scattering of electrons from interfaces, GBs and intrinsic defects in Cu$_2$O. Figure 8c shows the $\kappa_l$ as a function of T. It can be seen that $\kappa_l$ dominates over $\kappa_e$ in the whole T range. This is expected since transport of electrons through GBs and interface is difficult while phonons can carry heat via local phonon modes across the GBs and interface. $\kappa_l$ decreases as the fraction of Cu$_2$O increases in NC6 (figure 8c) and the trend is very similar to that of $\kappa$ (figure 8a). The low value of $\kappa_l$ is mainly attributed to multiscale phonon scattering from intrinsic points defects, interfaces and GBs along with softening of Raman active phonon mode $T_{2g}$ (figure 4b). The Cu/Cu$_2$O interface plays a central role in the reduction in $\kappa$ owing to large lattice mismatch of around 15 % (table S2) and dissimilar vibrational properties of Cu and Cu$_2$O[46].

For instance, Debye temperature difference of 207 K arises between Cu and Cu$_2$O as they are ~ 343 K and ~ 550 K, respectively[47,48]. Furthermore, their mass density ($\rho_m$) of 8.93 and 6.0 g/cm$^3$ with respective speed of sound (Vs) 2325 and 4687 m/s leads to respective acoustic impedance (Z = $\rho_m$Vs) of 2.076 and $2.812\times10^7$ kg.m$^{-2}$s$^{-1}$ (refs. [47,48]). Therefore, this larger acoustic impendence and lattice mismatch between Cu and Cu$_2$O and point defects, GBs and softening in phonon mode $T_{2g}$ favour low $\kappa_l$. Notably, the phonon dispersions of Cu and Cu$_2$O are significantly different since Cu$_2$O exhibits rich phonon DOS between 20 cm$^{-1}$ to 125 cm$^{-1}$ while Cu has very less phonon DOS in this frequency range near gamma point[18,48,49]. This will lower the communications among phonon modes, and will lead to drastic decrease in phonon transmission[41] and hence they are likely to decrease $\kappa$.

It is interesting to note that these trends of significantly reduced $\kappa$ and features are completely deviated from that of bulk Cu or Cu$_2$O[18]. The region of $\kappa$ in the range of 30 K to 100 K can be well-fitted by two different regimes (figure 8d). It follows the $T^x$ law, with x in the range of 0.37 to 0.79 in the low T regime and 0.32 to 0.50 in the high T regime (table S3). The power dependence of $\kappa$ on T decreases as Cu:Cu$_2$O decreases that is line with more intrinsic defects and interface in NC6. This is a significant departure from $T^3$ dependence on $\kappa$ of these individual components. Therefore, these results manifest dominant role of interface, defects and GBs as well as size effects, in consistent with earlier reports on metal/semiconductor interface[39] or semiconductor[44,44]. The value of $\kappa \sim 0.49$ W/m-K at 300 K in NC6 is significantly reduced compared to either Cu or Cu$_2$O, but comparable to quantum dot superlattices of PbTe and multi-element alloys at 300 K[34,50]. The $\kappa$ at 300 K of NC6 is reduced approximately by an order of magnitude to three orders of magnitude compared to those of the bulk $Cu_2O$ and copper metal. Furthermore, $\kappa \sim 0.14$ W/m-K at 10 K in NC6 is incredibly low, just six times that of air at 273 K and 1 atm. We attribute this to scattering of low-mid and high wavelength phonons from defects including point defects, vacancies, dislocations, GBs and mismatched interfaces as shown in figure 8e.

### 3.2.4 Power factor ($S^2\sigma$) and figure of merit (ZT)

There is evolution of power factor $S^2\sigma$ with increasing $Cu_2O$ fraction (figure 9a). Two dip-like sharp features near 41 K and 96 K for NC1, and 34 K and 93 K for NC2 are due to the sign crossover from negative to positive and from positive to negative in S. The position of dip decreases with increase in $Cu_2O$ fraction. However, only one dip each at 19 K, 66 K, 95 K, and 135 K is respectively found in NC3, NC4, NC5, and NC6. It increases with increase in fraction of $Cu_2O$. Its value at 300 K increases from NC1 to NC5 and then decreases in NC6, it initially decreases from NC1 to NC2 then increases at 10 K, with decrease in $Cu:Cu_2O$ fraction (figure 9a, inset).

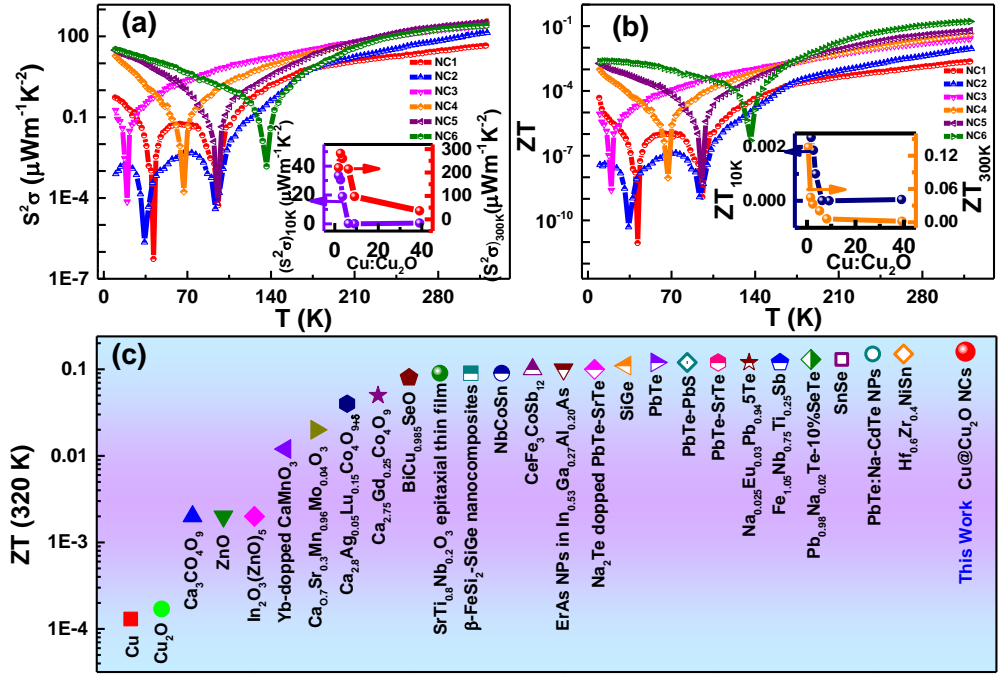

**Figure 9** (a) Power factor and (b) figure of merit ZT of NC1 to NC6. Insets a & b show power factor and ZT as a function of $Cu:Cu_2O$, respectively. (c) ZTs of various compounds that show the present result is significantly large, the largest.

Variation in ZT with T (figure 9b) is similar to that of $S^2\sigma$ (figure 9a). It increases with decrease in $Cu:Cu_2O$ at 10 K and 300 K (figure 9b, inset). Overall, it is the largest at 10 K or 300 K. The peak value of $S^2\sigma$ is found to be about 347 $\mu Wm^{-1}K^{-2}$ at 320 K in NC4 whereas NC6 ($Cu:Cu_2O=1.74$) exhibits maximum ZT of 0.16 at 320 K that is remarkably larger than those of Cu, $Cu_2O$ and several other TE materials (figure 9c). Overall, vacancies present in $Cu_2O$ along with percolation through Cu nanodomain networks lead to optimized electrical conductivity. Significantly, large value of S is due to filtering of low energy charge carriers along with sharp features in carrier DOS in these NCs, point defects in $Cu_2O$ and possible enhanced DOS near Fermi level due to charge transfer from Cu to $Cu_2O$. Ultralow κ is mainly due to multiscale phonon scattering of both low and high frequency phonons from point defects, formed interface, and GBs along with softening in Raman active mode in $Cu_2O$. They lead to high value of ZT (figure 9c), which indicates its potential for TE applications near room temperature with added advantages of nontoxic and low-cost components synthesized using a facile method.

## 4. Conclusion

We have identified the low-cost and nontoxic core-shell $Cu@Cu_2O$ nanocomposites with ultralow thermal conductivity ~ 0.49 $Wm^{-1}K^{-1}$, optimized electrical conductivity ~ 0.54 mΩm, and large thermopower ~ 373 $\mu VK^{-1}$, which are critical for advances in TE research. As a result, a significantly large ZT of around 0.16 at 320 K is obtained which is larger than those of many well-known TE materials such as PbTe, SnSe and SiGe, showing potential for thermoelectric cooling applications. We attribute these results to carrier energy filtering, sharp features in DOS, charge transfer among Cu and $Cu_2O$ interfaces, and multiscale phonon scattering due to formation of metal/semiconductor interfaces, vacancies, point defects, dislocations, and GBs. This study may thus open a new pathway in the development of new TE devices.

**Acknowledgment**

Authors would like to gratefully acknowledge Dr. D. M. Phase/G. Panchal and Dr. M. Gupta/L. Behera, Dr. V. Sathe, Dr. U. Deshpande, Ms. Pritee Bhardwaj, and Dr. S. C. Das, UGC-DAE Consortium for Scientific Research, Indore, India for providing XRD, Raman, FTIR, TEM data, and annealing facility, respectively, Hari Singh Gour University, Sagar for FESEM data and MNIT Jaipur EDX data.